\documentclass[preprint]{aastex}

\shorttitle{Close binary stars.~V}
\shortauthors{Rucinski \& et al.}

\begin{document}

\title{Radial Velocity Studies of Close Binary 
Stars.~V\footnote{Based on the data obtained at the David Dunlap 
Observatory, University of Toronto.}}

\author{Slavek M. Rucinski, Wenxian Lu\footnote{Current address:
  NASA/GSFC, Mailstop 664, 8800 Greenbelt Road, Greenbelt, MD 20771},
\ Stefan W. Mochnacki\footnote{Temporary address: Carnegie Observatories,
  813 Santa Barbara Street, Pasadena, CA 91101}}
\affil{David Dunlap Observatory, University of Toronto\\
P.O.Box 360, Richmond Hill, Ontario, Canada L4C 4Y6}
\email{rucinski,lu,mochnacki@astro.utoronto.ca}

\and

\author{Waldemar Og{\l}oza and Greg Stachowski}
\affil{Mt. Suhora Observatory of the Pedagogical University\\
ul.~Podchora\.{z}ych 2, 30-084 Cracow, Poland\\ and
N.Copernicus Astronomical Center\\
ul.~Bartycka 18, 00-716 Warszawa, Poland}
\email{ogloza,greg@wsp.krakow.pl}


\begin{abstract}

Radial-velocity measurements and sine-curve fits to the orbital 
velocity variations are presented for the fifth set of ten close binary 
systems: V376~And, EL~Aqr, EF~Boo, DN~Cam, FN~Cam, V776~Cas, SX~Crv,
V351~Peg, EQ~Tau, KZ~Vir. All systems are 
double-lined spectroscopic contact binaries (KZ~Vir may be
a low inclination, close, non-contact binary), with seven
(all except EL~Aqr, SX~Crv and EQ~Tau) being the recent
photometric discoveries of the Hipparcos satellite project. 
The most interesting object is SX~Crv, 
a contact system with an unprecedently low mass ratio, 
$q=0.066 \pm 0.003$, whose existence challenges
the current theory of tidal stability of contact systems. 
Several of the studied systems are prime candidates for combined 
light and radial-velocity synthesis solutions.
\end{abstract}


\section{INTRODUCTION}
\label{sec1}

This paper is a continuation of the radial-velocity studies of close 
binary stars, \citet[Paper I]{ddo1}, \citet[Paper II]{ddo2} and
\citet[Paper III]{ddo3}, \citet[Paper IV]{ddo4}.
The main goals and motivations are described in these papers.
In short, we attempt to obtain radial-velocity data for
close binary systems which are accessible to 1.8-m class 
telescopes at medium spectral resolution of about R = 10,000 -- 15,000. 
Selection of the objects is quasi-random in the sense that we have
started with short period (mostly contact) binaries and we
attempt to slowly progress to longer periods as the project
continues.
We publish the results in groups of ten systems as soon as reasonable
orbital elements are obtained from measurements evenly distributed 
in orbital phases. 
  
This paper is structured in the same way as Papers~I -- IV in that 
most of the data for the observed binaries are in 
two tables with the radial-velocity measurements 
(Table~\ref{tab1}) and their sine-curve solutions (Table~\ref{tab2}).
Section~\ref{sec2} of the paper contains brief summaries of 
previous studies for individual systems. 
The observations reported in this paper have been collected between
October 1996 and February 2001; the ranges of dates for
individual systems can be found in Table~\ref{tab1}.
All systems discussed in this paper have been observed for
radial-velocity variations for the first time. We have derived
the radial velocities in the same way as described in previous
papers. See in particular Paper~IV for a discussion of the
broadening-function approach used in the derivation of
the velocity amplitudes, $K_i$, and the center-of-mass
velocity, $V_0$.

The data in Table~\ref{tab2} are organized in the same manner as 
before. In addition to the parameters of spectroscopic orbits,
the table provides information about the relation between 
the spectroscopically observed epoch of the primary-eclipse T$_0$ 
and the recent photometric determinations in the form of the (O--C) 
deviations for the number of elapsed periods E. It also contains
our new spectral classifications of the program objects. 

For further technical details and conventions used in the paper, please
refer to Papers I -- IV of this series.

\placetable{tab1}

\placetable{tab2}

\section{RESULTS FOR INDIVIDUAL SYSTEMS}
\label{sec2}

\subsection{V376 And}

V376 And is one of the systems discovered by the Hipparcos satellite
mission \citep[HIP]{hip}. The early spectral type of the binary is
its distinguishing feature. Our estimate is A4V, not as early as the
previous estimates (A0), yet early enough to point out that otherwise
the system looks like a perfect contact binary of the W~UMa-type.
There are very few so early-type systems among W~UMa-type binaries.
The relatively long period, 0.799 days, is consistent with the
spectral type.

The light curve shows two equally deep minima, so that the choice
of the contact-binary type is somewhat arbitrary. We chose the
original \cite{hip} ephemeris which then leads to the A-type system
(the more massive star eclipsed at minimum corresponding to our $T_0$).
Note, however, that $T_0=$2,451,510.5416, determined by  
\cite{kes00} from photometric observations
obtained during the span of our spectroscopic observations, must 
then refer to the secondary minimum. These observations are 
in perfect agreement with our observations in terms of the initial
epoch, if allowance of a half-period is made.

The average $B-V = 0.28$ estimated from Tycho-2 project data
\citet[TYC2]{tyc2} agrees with our spectral type if 
some reddening of $E_{B-V} = 0.1$ is assumed. Then, 
the maximum-light 
magnitude\footnote{The maximum-light magnitude for this 
and subsequent stars has been estimated
in two ways, with the average value adopted: 
(1)~On the basis of maximum magnitude in HIP ($Hp$(5\%))
using a correction, $V-Hp=-0.03+0.155\,(B-V)$, which is applicable
in the range $0.1 < B-V < 0.8$, (2)~From the average
$V$ in TYC2, corrected by one-half of the photometric amplitude.}
$V_{max} \simeq 7.63$ and the HIP parallax 
$p=5.07 \pm 0.89$ milli-arcsec (mas) lead to $M_V = 0.85 \pm 0.40$.
The period--luminosity--color relation 
calibrated by \citet[RD97]{RD97}
gives $M_V = 1.1$. At the large distance of almost 200 pc, the
reddening of $E_{B-V} = 0.1$ is entirely possible. The system
is apparently one of the intrinsically
brightest and bluest contact binaries for which
the RD97 calibration appears to apply; the nominal range 
of the calibration is $0.26 < (B-V)_0 < 1.14$.

Our radial velocity data are of good quality (Figure~\ref{fig1}).
The system deserves a combined spectroscopic and
photometric analysis because of its rather
rarely observed combination of an early-type and of an apparently
very normal, unequal mass ($q=0.305 \pm 0.005$), 
equal-effective-temperature, contact binary. 

\placefigure{fig1}

\subsection{EL Aqr}

EL Aqr was discovered as a variable star by \citet{hof33} and then
re-discovered by \citet{ste68} who correctly identified it as
a contact binary system. He determined the period and gave
$V_{max} = 10.35$ and $B-V = 0.40$. While $V_{max}$ agrees 
estimates based on the HIP and TYC2 data, the latter suggest 
$B-V=0.55$, which would not agree well with our estimate of the
spectral type, F3V, unless the system suffers moderately large
reddening of $E_{B-V} \simeq 0.15$. The HIP parallax is small
and has large error, $p=4.7 \pm 1.9$, so that the absolute
magnitude is poorly determined.

Because of the relative faintness of EL Aqr and usually large
zenith distance as seen from the DDO, our radial-velocity
observations have relatively large scatter (Figure~\ref{fig1}. 
The system is of the A-type and has a small mass-ratio, 
$q=0.203 \pm 0.009$. The ephemeris of \citet{AH99} based on
photometric observations obtained during our observations
($T_0=$2,451,080.4443) agrees very well with our 
determination of $T_0$. 

\subsection{EF Boo}

EF Boo is one of the systems discovered by the Hipparcos satellite
mission \citep{hip}. Our solution is well defined 
(Figure~\ref{fig1}) and shows
a typical W-type system, albeit of a rather long orbital
period of 0.4205 days. The mass ratio is large, $q=0.512 \pm 0.008$,
as is common for W-type systems. The only peculiarity is the
spectral type which shows a complex spectrum consisting of a F5
component and of signatures of a late-G type. We have not been able
to detect a third component in the broadening functions, so that
the composite spectrum is a bit of a mystery. The mean $B-V=0.57$
from TYC2 is in fact consistent with a spectral type half way between
F5 and late-G. The HIP parallax $p=6.00 \pm 1.06$ mas is rather small,
but the resulting $M_V=3.15 \pm 0.40$, based on $V_{max}=9.26$ agrees
with the RD97 calibration, $M_V=3.47$, under an assumption
of no reddening.

EF Boo was observed photometrically during our observations
\citep{ozd01}. The closest primary eclipse time $T_0 =$2,451,719.330
agrees perfectly with our determination.

\subsection{DN Cam}

DN Cam is the next system discovered by the Hipparcos satellite
mission. It is similar to the one described
above, but it is by one magnitude
brighter, $V_{max}=8.14$. As the result, our 
observations show very little scatter (Figure~\ref{fig1}). 
The system appears to be an uncomplicated W-type
system, again of a rather long orbital period for W-type, 0.4983 days. 
Our spectral type, F2V, is based on only one spectrum, but it does
agree with the TYC2 average $B-V=0.36$ and no reddening. The parallax
is small, $p=4.49 \pm 0.89$ mas, so that the larger apparent
brightness than for EF~Boo must be due to a larger intrinsic luminosity.
From the parallax, $M_V=1.4 \pm 0.4$, while the RD97 calibration
gives $M_V(cal)=2.6$, so that there appears some discrepancy here.

DN Cam has not been observed photometrically since the discovery by
Hipparcos. Our determination of $T_0$ is fully consistent with
the original ephemeris, $T_0=$2,448,500.488. It should be noted that
the HIP light curve shows practically equally-deep eclipses, so
that the matter of the type of the system (A or W) is open and
awaits detailed modeling.

\placefigure{fig2}

\subsection{FN Cam}

FN Cam is another Hipparcos discovery. This time it is an A-type
system (Figure~\ref{fig2}) with a well defined radial-velocity
orbital solution and a moderately long period, 0.6771 days.
The original HIP epoch $T_0=$2,448,500.427 agrees well with our
determination of $T_0$. Again, the eclipses are almost equally
deep in this case.

Our independent estimates of the spectral 
type are not entirely consistent, A9 and
F2, but suggest at a definitely earlier spectral type than estimated
before (G2). The TYC2 value of
$B-V=0.41$ would be consistent with the spectral
type only if some reddening of about $E_{B-V}=0.05$
was present. For $V_{max}=8.47$, $p=3.62 \pm 0.95$ mas, 
$M_V=1.1 \pm 0.6$, while $M_V(cal)=2.0$.
FN~Cam has the largest value of $(M_1+M_2) \sin^3i = 
2.50 \pm 0.07 M_\odot$ among the binary systems of this group.

\subsection{V776 Cas}

V776~Cas was discovered by the Hipparcos mission. 
It appeared in \citet{due97} as V778~Cas with the 
period two times longer than the actual period of the binary.
The binary is a contact, A-type
system of a very small mass ratio, $q=0.130 \pm 0.004$. The 
HIP light curve has a small amplitude which is consistent with 
the relatively small $(M_1+M_2) \sin^3i = 0.975 \pm 0.026\,M_\odot$,
both most probably due to a low orbital inclination.

The star was observed photometrically after the HIP discovery by
\citet{gom99}. They did not publish their original eclipse times, but
reduced them to the HIP moments. The restored time 
for the time of their observations, as given in 
Table~\ref{tab2}, $T_0=$2,450,814.8957, very well agrees with our
determination.

Our spectral type, F2V, would agree with the TYC2 average 
$B-V=0.47$ if some reddening of about $E_{B-V}=0.1$ were
present. With $V_{max}=8.92$ and $p=4.86 \pm 1.56$, 
$M_V=2.0 \pm 0.7$, while the RD97 calibration predicts
$M_V(cal)=2.8$. 

V776 Cas is a brighter member of a visual binary ADS~1485.
The companion, at a separation of 5.38 arcsec, is 2 magnitude
fainter than the contact binary. We avoided the companion in our radial
velocity observations of V776~Cas, but observed its velocity on two
occasions, with the following results:
HJD = 2451769.800, $V_r=-26.4$ km/s and 
HJD = 2451806.715, $V_r=-27.4$ km/s.

\subsection{SX Crv}

This is definitely the most interesting among this group
of binaries because of the extremely small mass ratio
that we found,
$q=0.066 \pm 0.003$. This is not only a new ``record'',
surpassing the well-known case of AW~UMa with $q = 0.075$,
but also a bit of embarrassment for the theory which
currently predicts a cut-off at about
$q=0.09$ \citep{ras95}. The binary is similar to
AW~UMa in being the extreme mass-ratio A-type system,
but its spectral type is slightly later, F6/7V and
the period is shorter, 0.3166 days, compared with
F0/2V and 0.4387 for AW~UMa.

The discovery of spectral signatures of the secondary component
in SX~Crv was not easy. In fact, without any new photometric 
data, we unnecessarily collected many observations
during conjunctions, assigning our initial inability to detect
the secondary to a possible problem with the initial epoch
and/or to a variable period. Only later on did we realize that
a weak signature of the secondary was detectable 
in our broadening functions, in spite of the large ratio
of masses of 15:1. To calculate the orbital phases,
we used the HIP data on the period and the initial
epoch ($T_0=$2,448,500.1539). 

The binary was discovered by \citet{sun74}. Photometry
of \citet{sca84} showed a shallow light curve with an
amplitude of about 0.2 mag and $V_{max} \simeq 8.95$.
Our estimate is slightly fainter, $V_{max} = 8.99$. The
TYC2 color index, $B-V=0.52$, agrees with our spectral type 
F6/7V. The HIP parallax is the largest for this group of
binaries, $p=10.90 \pm 1.21$ mas leading to $M_V=4.18 \pm 0.25$,
which is consistent with the RD97 calibration, $M_V(cal) = 3.91$.

The HIP light curve is rather poorly covered so it is
difficult to say if the secondary eclipse of SX~Crv is total.
It may be actually the case because -- for a small $q$ --
total eclipses take place over a wide range of the inclinations;
also, an amplitude as ``large'' as the observed 0.2 mag.\
is not easy to obtain for such a small mass ratio 
without the inclination being sufficiently large
to produce total eclipses. $(M_1+M_2) \sin^3i =
0.861 \pm 0.028\,M_\odot$, when related to an expected mass
for the spectral type of F6/7V, suggests a moderately low 
inclination angle, $i \simeq 60 - 70$ degrees. Figure~3 in
\cite{MD72} gives the angle of the internal contact for
total eclipses. With $q=0.066$, no total secondary eclipses are
expected for $i < 66.5$ degrees, but for inclinations larger than
this angle the internal contact angle sensitively depends on $i$. 
For $i=70$ degrees and $q=0.066$, 
the internal contact angle already reaches a rather easy to
detect value of $\psi_i \simeq 14$ degrees. 
Thus, detection of totality in the secondary
eclipses will provide a powerful check on our spectroscopic
determination and a very good opportunity to determine
an accurate total mass of the system.

SX~Crv deserves a full combined solution based
on modern, accurate photometric observations. It is one of the most
interesting close binary systems currently known and
has a potential of increasing our knowledge of the critical,
perhaps final stages in contact, of very close binary systems.

\subsection{V351 Peg}

V351 Peg was discovered by the Hipparcos mission. It is listed
in the HIP catalog with the period equal one half of the actual
one (0.5933 days), apparently due to the identical depths 
of the eclipses. We used the period two times longer
than in the HIP catalogue, but
kept the original initial epoch for consistency
with these observations; then the system is of the W-type.
The system was subsequently photometrically observed by
\citet{gom99}. Their published value, advanced to the actual
time of the observations, is $T_0=$2,450,722.3918. Our
determination is fully consistent with this determination.

V351 Peg is a bright star, $V_{max}=7.92$, but its HIP parallax
is only moderately large, $p=7.34 \pm 0.92$ mas. The average
TYC2 color index 
$B-V=0.32$ requires some reddening of $E_{B-V}=0.05$ to
reconcile with our spectral type A8V
(the previous spectral type was A9III). With these data,
$M_V=2.09 \pm 0.28$, while $M_V(cal)=1.94$. Thus, the star
is not as blue and luminous as V376~And 
described at the beginning of this
section, but is nevertheless one of the good examples of
an early-type, genuine-contact, unequal component-mass system.

\placefigure{fig3}

\subsection{EQ Tau}

EQ Tau is the only system in this group which was not observed
by the Hipparcos satellite. It is best known for the fact that
it is located in the area of Pleiades and sometimes is mentioned
in this context \citep{PK84}. However a note in the General
Catalogue of Variable Stars \citep{gcvs} specifically mentions
that the system is not a member of Pleiades, probably on the
basis that it is too faint to be located on or above the
main sequence of the cluster.

The binary was discovered and observed by \citet{tse54}. New 
photometric observations were obtained by \citet{ben95}
and \citet{buc98}. We have used the more recent of these
eclipse time predictions, $T_0=$2,450,396.925 with $P=0.3413471$
days, which served our spectroscopic observations very well.

EQ Tau is too faint to be included in the HIP catalog, but
appears in TYC2 with the average $V=11.28$, $B-V=0.98$. The
color index is much redder than implied by our spectral type,
G2, although this estimate is uncertain. Possibly the star
is reddened by the interstellar matter in 
the Pleiades cluster. The proper motion of the star is 
exceptionally large, in fact the largest among stars of this
group, $\mu_\alpha\cos(\delta)=+68.4 \pm 1.4$ mas, 
$\mu_\delta=-29.6 \pm 1.5$ mas. If we assume
that the star is at the distance of Pleiades and use 
the Pleiades parallax following \citet{VanL99}, $p=8.45 \pm 0.25$
mas (118 pc), then $V_\alpha=+38$ km/s, $V_\delta=-17$ km/s
and $M_V \simeq 5.5$. For an assumed larger, 
arbitrary, but plausible distance of 200 pc 
($p=5$ mas), $V_\alpha=+65$ km/s, $V_\delta=-28$ km/s
and $M_V \simeq 4.3$. The large spatial velocity is certainly
strongly indicated by our  center-of-mass velocity, $V_0=+72$
km/s. Thus, the star, even if not a member of Pleiades, is
very interesting as a large spatial-motion object.
Despite relative faintness of EQ Tau for our telescope,
its spectroscopic orbit is relatively well defined 
(Figure~\ref{fig3}). 

\subsection{KZ Vir}

KZ Vir is another HIP discovery which, apparently, 
has not been observed photometrically since; for that
reason we used $T_0=$2,448,500.3165 from the HIP catalog.
KZ~Vir has been included in our program mostly because of its
large apparent brightness ($V_{max}=8.36$). It is an interesting 
object: If it is a contact binary, then it has a
surprisingly late spectral type, 
F7, for its relatively long orbital period of 1.13 days. 
The mass ratio is not typical for majority of contact binaries,
$q = 0.85$, but such rare systems occasionally exist.

The average TYC2 value of $B-V=0.48$ agrees with the spectral type
of F7. The HIP parallax, $p=6.53 \pm 0.93$, with assumed
no reddening, gives $M_V=2.43 \pm 0.32$. This luminosity
estimate is in some conflict with the RD97 
contact-binary calibration predicting 
a much larger brightness, $M_V(cal)=1.33$. This is a strong
indication that the system has much less radiating area
than a contact system and is a close, but detached binary.

The velocity amplitudes are relatively small, which would be 
consistent with the small photometric amplitude (0.07 mag.), both
probably caused by a low orbital inclination. The small
photometric amplitude could also result from a moderate
distortion of its detached components.

\section{SUMMARY}

The paper presents radial-velocity data and orbital solutions 
for the fifth group of ten close binary systems 
that we observed at the David Dunlap Observatory. 
None was observed spectroscopically before.
All systems are double-lined (SB2) binaries with 
visible spectral lines of both components. All, with
the possible exception of KZ~Vir with the period
of 1.13 days, are contact binaries. The values 
of $(M_1+M_2) \sin^3i = 1.0385 \times 10^{-7}\,(K_1+K_2)^3\,
P({\rm day})\,M_\odot$ are given in Table~\ref{tab2}. 

Although our selection of targets is quite unsystematic,
we keep discovering very interesting objects. 
Seven of the systems are new photometric discoveries of the
Hipparcos project. The magnitude-limited nature of the HIP 
survey has led to an emphasis on relatively luminous,
massive, early-type (middle-A to early-F) systems. 
V376~And, DN~Cam, FN~Cam, V776~Cas, V351~Peg are such
binaries; they have typical properties of contact 
systems, with the mass ratio far from unity.

The most interesting discovery of this series is undoubtedly
the very small mass ratio of SX~Crv, $q=0.066 \pm 0.003$. This
becomes now the smallest known mass ratio among contact binaries,
taking the record from the well-known AW~UMa.
The detection of the low-mass companion was possible mostly
thanks to the superior capabilities of the linear SVD 
broadening-function approach \citep{ruc99}.
Together with $q=0.970 \pm 0.003$ determined in Paper~III
for V753~Mon, our spectroscopic observations have resulted
in a widest opening up of the mass-ratio domain of contact
binaries. The very
small mass ratio for SX~Crv strongly suggests that AW~UMa
is not an exception and that mass ratios smaller than the
theoretically predicted lower limit of $q_{min} \simeq 0.09$
\citep{ras95} certainly exist. 
 
\acknowledgements
The authors would like to thank Jim Thomson, Mel Blake, Anita Chung,
Chun Du and Chris Capobianco for participating in our
program and Frank Fekel for providing us a list of 
early-type standard stars which we frequently use.

Support from the Natural Sciences and Engineering Council of Canada
to SMR is acknowledged with gratitude. 

The research has made use of the SIMBAD database, operated at the CDS, 
Strasbourg, France and accessible through the Canadian 
Astronomy Data Centre, which is operated by the Herzberg Institute of 
Astrophysics, National Research Council of Canada.

\clearpage

\noindent
Captions to figures:

\bigskip

\figcaption[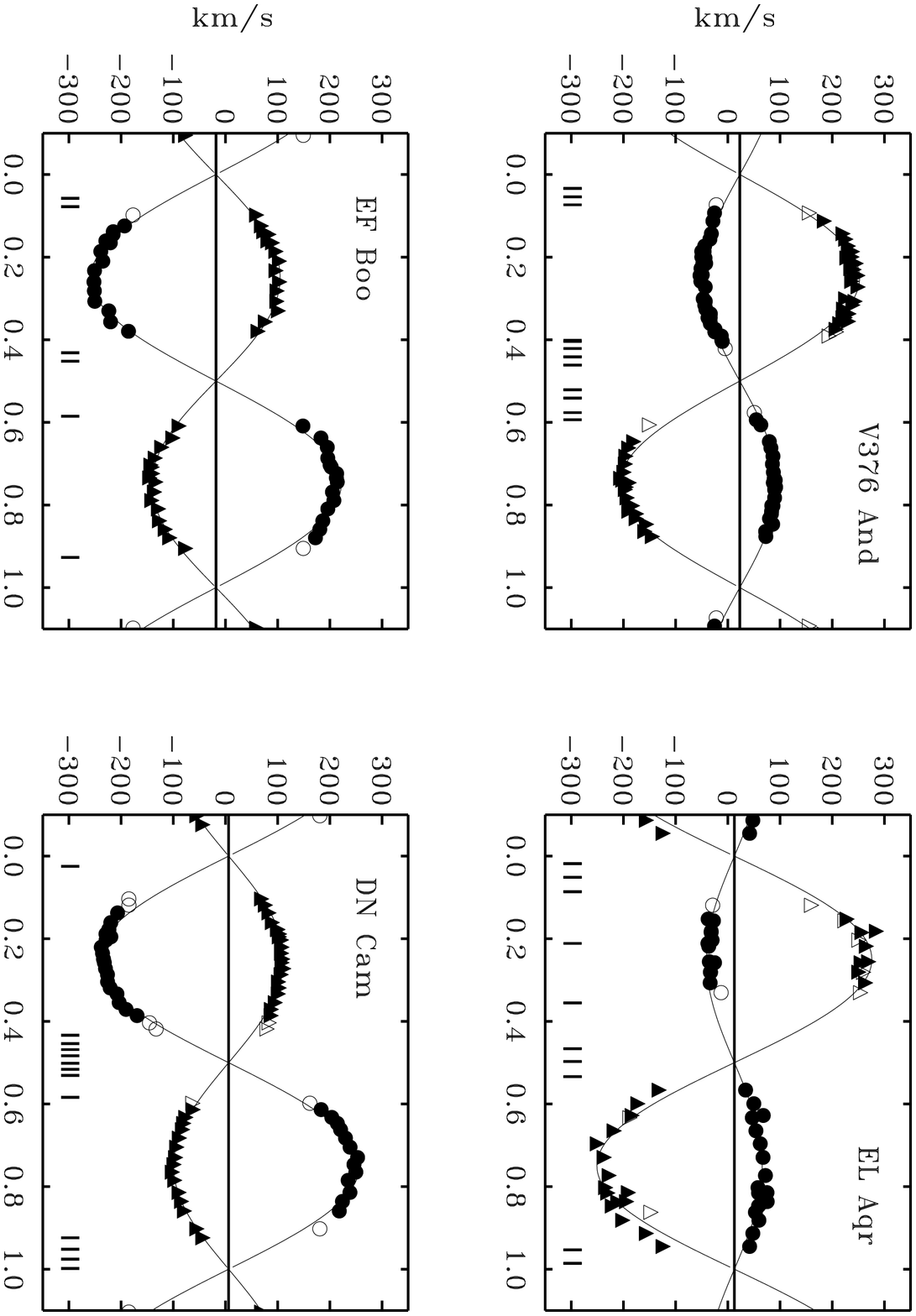] {\label{fig1}
Radial velocities of the systems V376~And, EL~Aqr, EF~Boo and
DN Cam are plotted in individual panels versus orbital phases. 
The lines give the respective circular-orbit (sine-curve) 
fits to the radial velocities. V376~And and EL~Aqr
are A-type systems while EF~Boo and DN Cam 
are W-type systems. The circles and triangles 
in this and the next two figures
correspond respectively to components eclipsed at $T_0$ 
(as given in Table~\ref{tab2}) or half a period later, 
while open symbols indicate observations contributing 
half-weight data in the solutions. Short marks in the lower 
parts of the panels show phases of available 
observations which were not used in the solutions because of the 
blending of lines. All panels have the same vertical 
ranges, $-350$ to $+350$ km\,s$^{-1}$. 
}

\figcaption[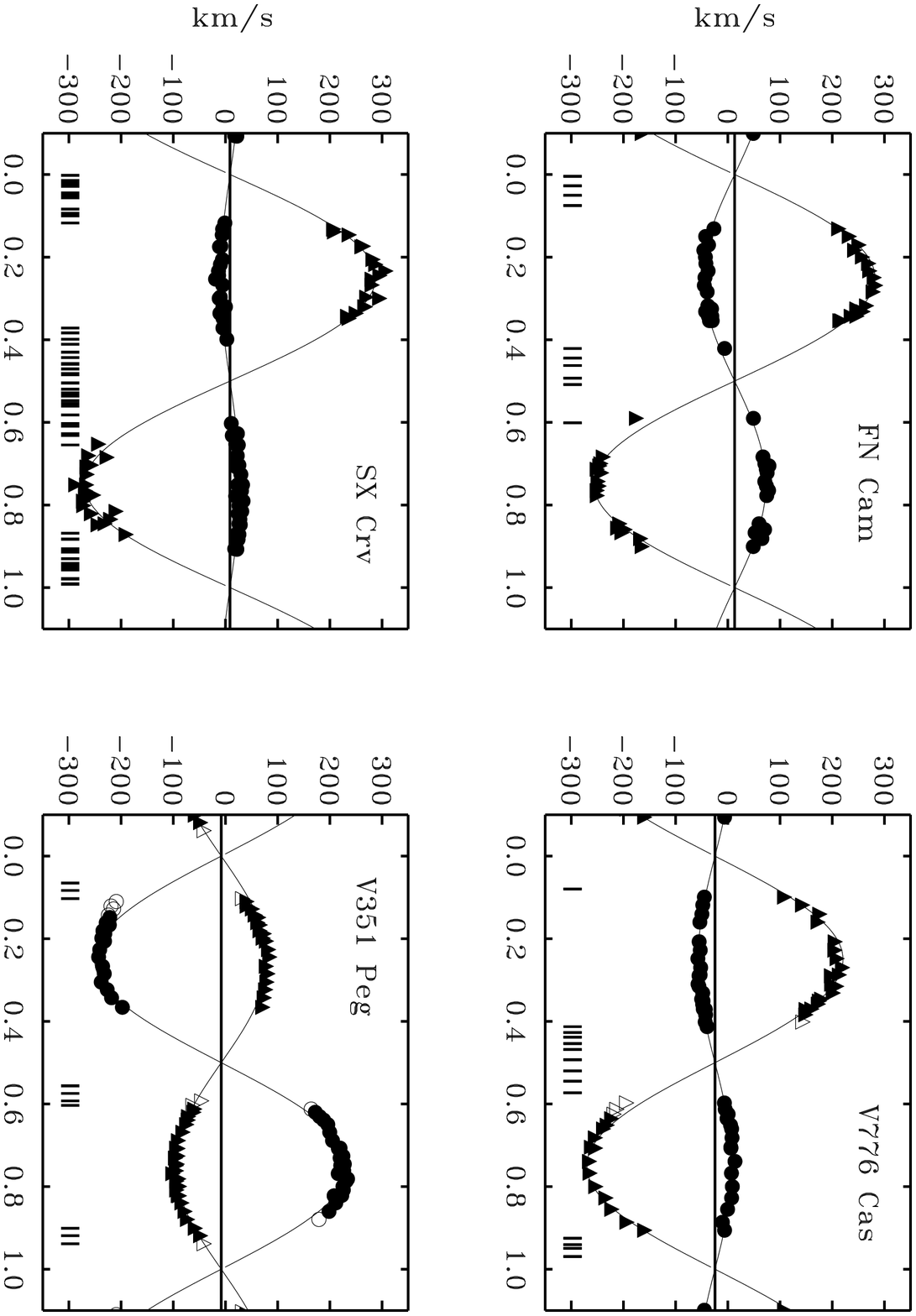] {\label{fig2}
Same as for Figure~1, but with the radial velocity orbits
for the systems FN~Cam, V776~Cas, SX~Crv and V351~Peg.
Only V351~Peg is a W-type system, the other are A-type
systems. SX Crv is a new record holder with the
mass ratio of only $q=0.066 \pm 0.003$. 
}

\figcaption[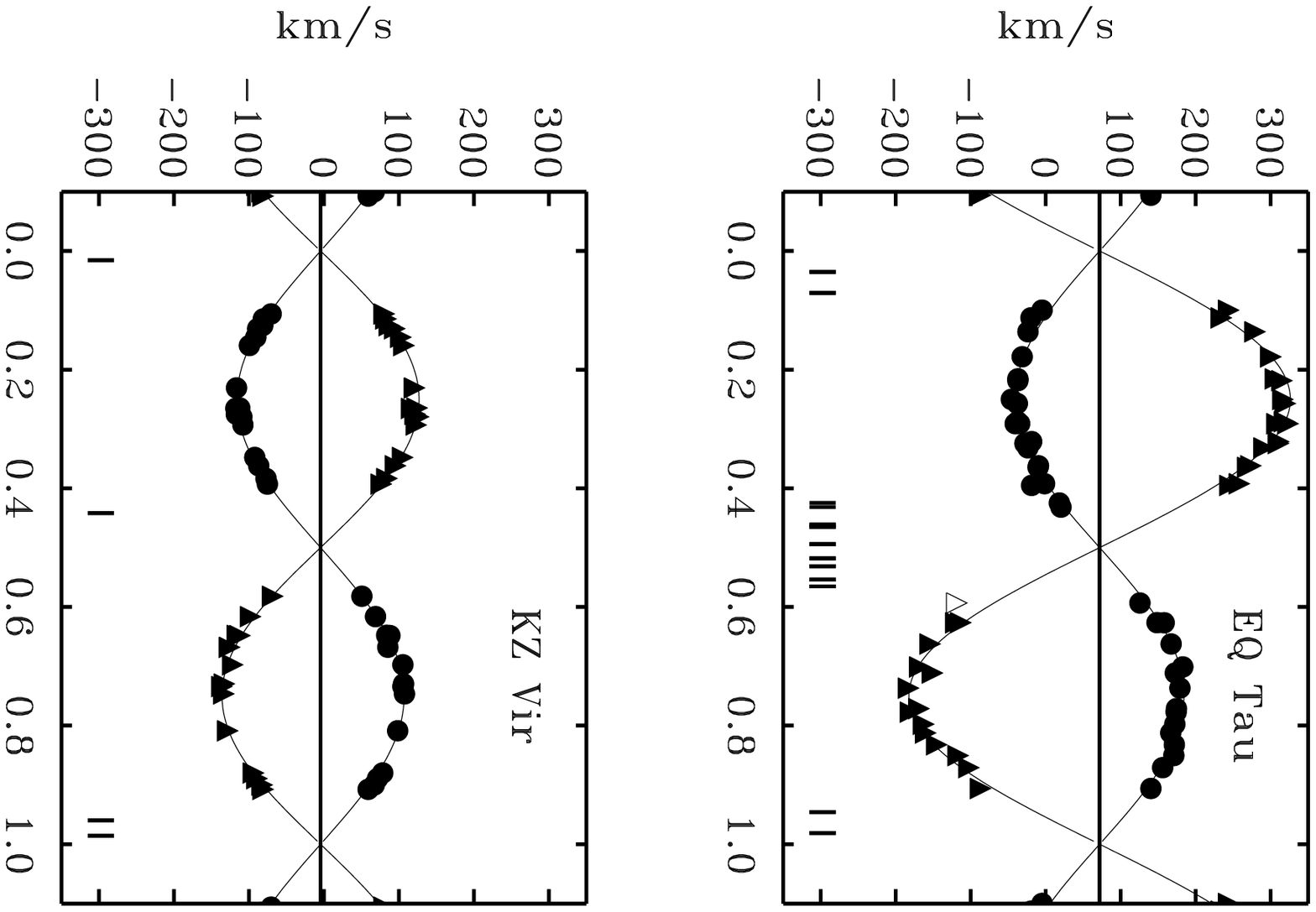] {\label{fig3}
Same as for Figure~1, for the systems EQ~Tau and KZ~Vir. 
EQ~Tau is a relatively inconspicuous system
close in the sky to Pleiades. It is most likely
behind the cluster and appears to have an
exceptionally large spatial velocity. This is indicated also
by the large center-of-mass velocity, $V_0$, 
as shown in the figure. KZ~Vir
is probably a close, but non-contact system of two similar,
distorted stars.
}

\clearpage                    

\begin{table}                 
\dummytable \label{tab1}      
\end{table}
 
\begin{table}                 
\dummytable \label{tab2}      
\end{table}


\begin{thebibliography}{}

\bibitem[Agerer \& Huebscher(1999)]{AH99}          
    Agerer, F., \& Huebscher, J. 1999, 
    Inf.\ Bull.\ Variable Stars, 4711
\bibitem[Benbow \& Mutel(1995)]{ben95}             
    Benbow, W. R. \& Mutel, R. L.
    1995, Inf.\ Bull.\ Variable Stars, 4187
\bibitem[Buckner et al.(1998)]{buc98}          
    Buckner, M., Nellermore, B. \& Mutel, R. L.
    1998, Inf.\ Bull.\ Variable Stars, 4559 
\bibitem[Duerbeck(1997)]{due97}                  
    Duerbeck, H. W.
    1997, Inf.\ Bull.\ Variable Stars, No.4513
\bibitem[ESA(1997)]{hip}                           
     European Space Agency. 1997. The Hipparcos and Tycho
     Catalogues (ESA SP-1200)(Noor\-dwijk: ESA) (HIP)
\bibitem[Gomez--Forrellad et al.(1999)]{gom99}   
    Gomez--Forrellad, J. M., Garcia--Melendo, E., Guarro--Flo, J., 
    Nomen--Torres, J. \& Vidal--Saintz, J.
    1999, IBVS, 4702
\bibitem[Hoffmeister(1933)]{hof33}               
    Hoffmeister, C.
    1933, AN, 247, 281
\bibitem[Hog(2000)]{tyc2}                        
    Hog, E., Fabricius, C., Makarov, V. V., Urban, S., Corbin, T.,
    Wycoff, G., Bastian, U., Schwekendiek, P. \& Wicenec, A.
    2000, \aap, 355, L27 (TYC2)
\bibitem[Keskin et al.(2000)]{kes00}             
    Keskin, V., Yasarsoy, B. \& Sipahi, E.
    2000, IBVS, 4855
\bibitem[Kholopov et al.(1985--1988)]{gcvs}
    Kholopov, P. N., Samus, N. N., Frolov, M. S.
    Goranskij, V. P., Gorynya, N. A., Karitskaja, E. A., 
    Kazarovets, E. V., Kireeva, N. N., Kukarkina, N. P.,
    Kurochkin, N. E.,  Medvedeva, G. I., Pastukhova, E. N.,
    Perova, N. B., Rastorguev, A. S. \& Shugarov, S. Yu.
    1985--1988, General Catalogue of Variable Stars, 4th Edition
    (Moscow: Nauka Publishing House)
\bibitem[van Leeuwen(1999)]{VanL99}              
    van Leeuwen, F.
    1999, \aap, 341, L71
\bibitem[Lu \& Rucinski(1999)]{ddo1}             
    Lu, W., \& Rucinski, S. M. 
    1999, \aj, 118, 515 (Paper I)
\bibitem[Lu et al.(2001)]{ddo4}                  
    Lu, W., Rucinski, S. M. \& Ogloza, W. 
    2001, \aj, in press, AJ July 2001 (Paper IV)
\bibitem[Mochnacki \& Doughty(1972)]{MD72}
    Mochnacki, S. W. \& Doughty, N. A.
    1972, \mnras, 156, 51
\bibitem[Ozdemir et al.(2001)]{ozd01}            
    Ozdemir, S., Demircan, O., Erdem, A.,
    Cicek, C., Bulut, I., Soydugan, F. \& 
    Soydugan, E.
    2001, Inf.\ Bull.\ Variable Stars, 5033
\bibitem[Popova \& Kraicheva(1984)]{PK84}        
    Popova, M. \& Kraicheva, Z.
    1984, AISAO (=Astr. Isl. Spec. Astroph. Obs.), 18, 64
\bibitem[Rasio(1995)]{ras95}                     
    Rasio, F. A.
    1995, \apj, 444, L41
\bibitem[Rucinski(1999)]{ruc99}                  
    Rucinski, S. M.
    1999, ``Precise Stellar Radial Velocities'', 
    ASP Conf.\ Ser.\ Vol.185, 
    eds. J. B. Hearnshaw \& C. D. Scarfe, p.82
\bibitem[Rucinski \& Duerbeck(1997)]{RD97}       
     Rucinski, S.M., \& Duerbeck, H.W. 1997, 
     \pasp, 109, 1340  (RD97)             
\bibitem[Rucinski \& Lu(1999)]{ddo2}               
    Rucinski, S. M. \& Lu, W.
    1999, \aj, 118, 2451 (Paper II)
\bibitem[Rucinski et al.(2000)]{ddo3}              
     Rucinski, S. M., Lu, W. \& Mochnacki, S. W.
     2000, \aj, 120, 1133 (Paper III)
\bibitem[Scaltriti \& Busso(1984)]{sca84}          
     Scaltriti, F. \& Busso, M.
     1984, \aap, 135, 23
\bibitem[Stepien(1968)]{ste68}                    
     Stepien, K.
     1968, \pasp, 80, 220                         
\bibitem[Sunwal et al.(1974)]{sun74}              
     Sunwal, N. B., Sarma, M. B. K., Parthasarathy, M. \&
     Abhyankar, K. D. 
     1974, \aaps, 13, 81
\bibitem[Tsesevich(1954)]{tse54}                  
     Tsesevich, B. P. 
     1954, Izv.\ Astr.\ Obs.\ Odessa, 4, No.3

\end{thebibliography}
\end{document}


\begin{deluxetable}{lccrrrccc}

\tabletypesize{\scriptsize}

\pagestyle{empty}
\tablecolumns{9}

\tablewidth{0pt}
\tablenum{2}
\tablecaption{Spectroscopic orbital elements of the fifth
ten of close binary systems}
\tablehead{
   \colhead{Name} &                
   \colhead{Type} &                
   \colhead{Other names} &         
   \colhead{$V_0$} &               
   \colhead{K$_1$} &               
   \colhead{$\epsilon_1$} &        
   \colhead{T$_0$ -- 2,400,000} &  
   \colhead{P (days)} &            
   \colhead{$q$}          \\       
   \colhead{}     &                
   \colhead{Sp.~type}    &         
   \colhead{}      &               
   \colhead{} &                    
   \colhead{K$_2$} &               
   \colhead{$\epsilon_2$} &        
   \colhead{(O--C), [E]} &         
   \colhead{$(M_1+M_2) \sin^3i$} &    
   \colhead{}                      
}
\startdata

V376 And & EW/A  & HD~15922   &  +22.83(0.89) & 70.00(0.67) &
 4.17  & 51,644.3215(24) & 0.798669         & 0.305(5) \\
       & A4V     & HIP~12039  &             & 229.67(1.78) &
 13.20 & $+0.0028$ [167.5] & 2.232(55) &             \\[1mm]

EL Aqr & EW/A    & BD$-08^\circ 6189$ &  +12.51(2.08) & 53.38(1.53) &
 8.88  & 51,109.3334(32) & 0.481410         & 0.203(8) \\
       & F3V     & HIP~117317 &             & 263.34(4.38) &
23.49  & $+0.0049$ [60]    & 1.588(89) &             \\[1mm]

EF Boo & EW/W    & HD~234150   & $-13.16$(1.07) & 122.84(1.10) &
 5.28  & 51,725.2174(12) & 0.429512         & 0.512(8) \\
       & F5+lateG:   & HIP~71107 &          & 239.90(2.25) &
11.67  & $+0.0003$ [14]  & 2.129(59) &                  \\[1mm]

DN Cam & EW/W    & HD~29213   & +6.04(0.98) & 105.45(0.65) &
 4.13  & 51,679.6954(12) & 0.498312         & 0.421(6) \\
       & F2V:    & HIP~21913 &              & 250.62(1.91) &
13.04  & $-0.0232$ [6380]  & 2.336(50) &             \\[1mm]

FN Cam & EW/A    & HD~79886   & +12.96(1.11) & 59.62(0.93) &
 5.20  & 51,351.1554(24) & 0.677128         & 0.222(5) \\
       & A9, F2   & HIP~46005 &              & 269.01(2.10) &
11.50  & $+0.0195$ [4210]  & 2.496(69) &             \\[1mm]

V776 Cas & EW/A  & BD+69$^\circ$121 & $-24.71$(0.69) & 31.97(0.64) &
 2.58  & 51,788.2184(8) & 0.440413          & 0.130(4) \\
       & F2V   & HIP~8821 &                  & 245.31(1.83) &
12.04  & $+0.0099$ [2210]\tablenotemark{a} & 0.975(26) &       \\[1mm]

SX Crv & EW/A  & HD~110139     & +8.71(0.94) &    18.28(0.74) &
 3.98  & 51,070.8192(10) & 0.316622          & 0.066(3) \\
       & F6V, F7V   & HIP~61825 &                 & 278.70(2.43) &
21.44  & $+0.0129$ [8119]   & 0.861(28) &            \\[1mm]

V351 Peg & EW/W  & HD~220659     & $-8.08$(0.89) &   87.37(0.83) &
 3.83  & 51,418.3205(15) & 0.593297            & 0.360(6) \\
       & A8V   & HIP~115627 &               & 242.62(1.99) &
16.89  & $-0.0091$ [1173]\tablenotemark{b}  & 2.214(57) &            \\[1mm]

EQ Tau & EW/A  &   \nodata       & +71.95(1.22) &  112.41(1.43) &
 7.46  & 51,183.9030(9) & 0.341348          & 0.442(7) \\
       & G2:   &   \nodata      &               & 254.38(2.42) &
13.15  & $+0.0023$ [2305]   & 1.749(55) &        \\[1mm]

KZ Vir & EW/A  & HD~114726       & $-4.50$(0.48) &  111.40(0.77) &
 3.10  & 51,414.7848(11) & 1.131820          & 0.848(8) \\
       & F7V   & HIP~64433  &               & 131.34(0.87) &
 3.82  & $+0.0318$ [2575]   & 1.681(34) &        \\[1mm]

\tablenotetext{a}{V776 Cas: $T_0$ has been advanced from the
published time to the mean time of observations: HJD = 2,450,814.8957.}

\tablenotetext{b}{V351 Peg: Same as for V776 Cas: HJD = 2,450,722.3918.}

\tablecomments{The spectral types given in column two
are all new. The convention of naming the
binary components is that the subscript 1 designates
the component which is more massive so that
the mass ratio is defined to be always $q \le 1$.
The standard errors of the circular 
solutions in the table are expressed in units of last decimal places 
quoted; they are given in parantheses after each value. 
For example, the last table entry for the mass ratio 
$q$, 0.848(8), should be interpretted as $q = 0.848 \pm 0.008$.
The center-of-mass velocities ($V_0$), the 
velocity amplitudes (K$_i$) and the standard unit-weight 
errors of the solutions ($\epsilon$) are expressed 
in km~s$^{-1}$. The spectroscopically determined 
moments of primary minima are given by $T_0$; the corresponding
(O--C) deviations (in days) have been calculated from the most 
recent available ephemerides, as given in the text,
using the assumed periods and the number of epochs given by [E].
The values of $(M_1+M_2)\sin^3i$ are in the solar mass units.
}
\enddata
\end{deluxetable}